\documentclass{ws-ijmpa}

\begin{document}

\markboth{G.~L.~Klimchitskaya, E.~V.~Blagov \& V. M. Mostepanenko}
{Problems in the Lifshitz Theory of Atom-Wall Interaction}

%
\catchline{}{}{}{}{}
%

\title{PROBLEMS IN THE LIFSHITZ THEORY OF ATOM-WALL INTERACTION}

\author{G. L. KLIMCHITSKAYA\footnote{On leave from
North-West Technical University, St.Petersburg, Russia}}

\address{Institute for Theoretical Physics, Leipzig University,
D-04009, Leipzig, Germany \\
Galina.Klimchitskaya@itp.uni-leipzig.de}

\author{E. V. BLAGOV}

\address{Noncommercial Partnership
``Scientific Instruments'',  Moscow,  Russia}

\author{V. M. MOSTEPANENKO\footnote{On leave from Noncommercial Partnership
``Scientific Instruments'',  Moscow,  Russia}}

\address{Institute for Theoretical Physics, Leipzig University,
D-04009, Leipzig, Germany}

\maketitle

\begin{history}
\received{30 October 2008}
\revised{12 January 2009}
\end{history}

\begin{abstract}
Problems in the Lifshitz theory of atom-wall interaction arise when
the dc conductivity of dielectric wall is included into the model
of the dielectric response. We review the low-temperature behavior
of the free energy and entropy of dispersion interaction
for both dielectric and
metallic walls. Consistency of the obtained results with thermodynamics
and experimental data is analyzed. Recent attempts to include the
screening effects and diffusion currents into the Lifshitz theory are
considered. It is shown that this leads to the violation of the Nernst
heat theorem for wide classes of wall materials. The physical reasons for
the emergence of thermodynamic and experimental inconsistencies are
elucidated.

\end{abstract}

\ccode{PACS numbers: 12.20.-m, 34.35.+a, 78.20.Ci}

\section{Introduction}
The Lifshitz theory describes dispersion interaction between an atom or a
molecule and a cavity wall which is caused by quantum fluctuations of
the electromagnetic field. At separations from a few angstroms to a few
nanometers the interaction of an atom with a wall is of nonrelativistic
character. In this separation region it is described by the nonretarded
van der Waals potential. At larger separations the relativistic effects
come into play. At separations of about $1\,\mu$m  the atom-wall
interaction is described by the Casimir-Polder potential.
The dispersion interaction of atoms and molecules with walls made of
different materials plays important role in physical, chemical and
biological processes.\cite{1} During the last few years special attention
was attracted to the role of atom-wall interaction in Bose-Einstein
condensation and quantum reflection (see, e.g.,
Refs.~\refcite{2}--\refcite{6} and references therein).
In the framework of the Lifshitz theory material properties of a wall are
described by the frequency-dependent dielectric permittivity and the
properties of an atom by the atomic dynamic polarizability.

It is well known that the Lifshitz theory faces problems when the drift
current of conduction electrons is taken into account in the model of the
dielectric response (see the review in Ref.~\refcite{7}). These problems,
connected with the violation of the third law of thermodynamics
(the Nernst heat theorem) and contradictions between the theoretical
predictions and experimental data, where mostly discussed in the case
of two material plates. Here, we consider the complicated problems
arising in the Lifshitz theory of atom-wall interaction which are connected
with the description of free charge carrirs in the wall material.
In Sec.~2 we review known results for the Casimir-Polder free energy and
entropy of atom-wall interaction in the limit of low temperatures.
We show that for metallic walls theory is thermodynamically consistent,
whereas for dielectric walls it is consistent with thermodynamics only if
the dc conductivity of wall material is neglected in the model of the
dielectric response. The physical reasons explaining why the difficulties
with thermodynamics emerge are elucidated.
In Sec.~3 the experimental test for the influence of charge carriers on the
Casimir-Polder force is considered. Section 4 presents recent
attempts\cite{8}\cdash\cite{10} to generalize the Lifshitz theory of the
Casimir-Polder interaction through the inclusion of screening effects and
diffusion currents. Here, we demonstrate that the obtained generalized
reflection coefficients lead to contradiction between the Lifshitz theory
and thermodynamics for a wide class of dielectric materials.
The physical reasons why the screening effects are irrelevant  to the
van der Waals and Casimir-Polder interaction are explained. In Sec.~5
the reader will find our conclusions and discussion.

\section{Asymptotic Behavior of the Casimir-Polder Free Energy and
Entropy at Low Temperatures}

The free energy of dispersion interaction of an atom (molecule)
separated by a distance $a$ from a plate at temperature $T$ in thermal
equilibrium with environment is given by the Lifshitz
formula\cite{2}\cdash\cite{4,11}
\begin{equation}
{\cal F}(a,T)=-\frac{k_BT}{8a^3}\sum_{l=0}^{\infty}
{\vphantom{\sum}}^{\prime}\!\!\alpha({\rm i}\omega_c\zeta_l)
\int_{\zeta_l}^{\infty}\!\!\!\!dy{\rm e}^{-y}\left[(2y^2-\zeta_l^2)
r_{\rm TM}({\rm i}\zeta_l,y)-\zeta_l^2r_{\rm TE}({\rm i}\zeta_l,y)
\right].
\label{eq1}
\end{equation}
\noindent
Here, $\alpha(\omega)$ is the dynamic polarizability of an atom (molecule),
$\zeta_l=4\pi k_BTal/(\hbar c)$ [$l=0,\,1,\,2,\,\ldots\,$,
$k_B$ is the Boltzmann constant, $\omega_c=c/(2a)$] are the dimensionless
Matsubara frequencies, and the reflection coefficients for two independent
polarizations of the electromagnetic field (transverse magnetic and
transverse electric) are given by
\begin{equation}
r_{\rm TM}({\rm i}\zeta_l,y)=\frac{\varepsilon_ly-
\sqrt{y^2+\zeta_l^2(\varepsilon_l-1)}}{\varepsilon_ly+
\sqrt{y^2+\zeta_l^2(\varepsilon_l-1)}},
\qquad
r_{\rm TE}({\rm i}\zeta_l,y)=\frac{y-
\sqrt{y^2+\zeta_l^2(\varepsilon_l-1)}}{y+
\sqrt{y^2+\zeta_l^2(\varepsilon_l-1)}},
\label{eq2}
\end{equation}
\noindent
where $\varepsilon_l\equiv\varepsilon({\rm i}\omega_c\zeta_l)$ is the
dielectric permittivity of plate material along the imaginary frequency
axis. The prime adds a multiple 1/2 to the term of (\ref{eq1})
with $l=0$.

For dielectrics with neglected dc conductivity the dielectric permittivity
along the imaginary frequency axis is presented in the form\cite{1}
\begin{equation}
\varepsilon_l=1+\sum_{j}
\frac{g_j}{\omega_j^2+\omega_c^2\zeta_l^2+\gamma_j\omega_c\zeta_l}
\approx 1+\sum_{j}
\frac{g_j}{\omega_j^2+\omega_c^2\zeta_l^2},
\label{eq3}
\end{equation}
\noindent
where $\omega_j\neq 0$ are the oscillator frequencies, $g_j$ are the
oscillator strengths and $\gamma_j$ are the relaxation parameters.
This representation assumes that the static permittivity of a dielectric
material is finite:
\begin{equation}
\varepsilon_0=\varepsilon(0)=1+\sum_{j}
\frac{g_j}{\omega_j^2}<\infty.
\label{eq4}
\end{equation}
\noindent
The atomic dynamic polarizability can be represented with sufficient
precision using the single-oscillator model
\begin{equation}
\alpha({\rm i}\omega_c\zeta_l)=\frac{\alpha(0)}{1+\beta^2\zeta_l^2}
\label{eq4a}
\end{equation}
\noindent
with a dimensionless constant $\beta$.
The asymptotic behavior of the free energy (\ref{eq1}) combined with the
dielectric permittivity (\ref{eq3}) at low temperatures was found in
Ref.~\refcite{12}. Under the condition $\tau=2\pi T/T_{\rm eff}\ll 1$,
where the effective temperature is defined by $k_BT_{\rm eff}=\hbar\omega_c$,
it holds
\begin{equation}
{\cal F}(a,T)\approx E(a)-\frac{\hbar c\pi^3}{240a^4}\alpha(0)
C(\varepsilon_0)\left(\frac{T}{T_{\rm eff}}\right)^4.
\label{eq5}
\end{equation}
\noindent
Here, $E(a)$ is the Casimir-Polder energy at zero temperature and
$C(\varepsilon_0)$ is the function which goes to zero when
$\varepsilon_0\to 1$ (see Ref.~\refcite{12} for the explicit form of it).
For commonly used dielectrics, such as SiO${}_2$ with $\varepsilon_0=3.81$
and Si with $\varepsilon_0=11.67$, $C(\varepsilon_0)$ is equal to 2.70
and 6.33, respectively. From (\ref{eq5}) the Casimir-Polder entropy
is given by
\begin{equation}
S(a,T)=-\frac{\partial{\cal F}(a,T)}{\partial T}\approx
\frac{\pi^3k_B}{30a^3}\alpha(0)
C(\varepsilon_0)\left(\frac{T}{T_{\rm eff}}\right)^3.
\label{eq6}
\end{equation}
\noindent
As is seen from this equation, the Casimir-Polder entropy goes to zero
when $T$ vanishes, i.e., the Lifshitz theory with the permittivity
(\ref{eq3}) satisfies the Nernst heat theorem.

The model of the dielectric response (\ref{eq3}) does not take into account,
however, that all dielectrics at nonzero temperature possess small but
physically real dc conductivity. For dielectric materials conductivity
vanishes with temperature exponentially fast:
\begin{equation}
\sigma({\rm i}\omega_c\zeta_l,T)\sim\exp\left(
-\frac{\Delta}{k_BT}\right)\to 0\quad\mbox{when}\quad T\to 0,
\label{eq7}
\end{equation}
\noindent
where the physical meaning of the coefficient $\Delta$ is different for
different classes of dielectrics. The dc conductivity is usually taken
into account by means of the Drude-like term
\begin{equation}
\tilde\varepsilon_l=\varepsilon_l+
\frac{4\pi\sigma({\rm i}\omega_c\zeta_l,T)}{\omega_c\zeta_l}.
\label{eq8}
\end{equation}
\noindent
Here,
\begin{equation}
\sigma({\rm i}\omega_c\zeta_l,T)=
\frac{\sigma(0,T)}{1+\frac{\omega_c\zeta_l}{\gamma}},
\label{eq9}
\end{equation}
\noindent
where $\sigma(0,T)$ is the dc conductivity and $\gamma$ is the relaxation
parameter of free electrons.

The substitution of the permittivity (\ref{eq8}) into the Lifshitz formula
(\ref{eq1}) leads to only negligible additions to all terms with $l\geq 1$.
These additions exponentially decay to zero with vanishing
temperature.\cite{16}\cdash\cite{18} However, the term with $l=0$ is
modified because according to (\ref{eq2})
$r_{\rm TM}(0,y)\equiv r_0=(\varepsilon_0-1)/(\varepsilon_0+1)$ is
replaced with $\tilde{r}_{\rm TM}(0,y)=1$. As a result, with dc conductivity
included the free energy at low temperature is given by\cite{12}
\begin{equation}
\tilde{\cal F}(a,T)\approx{\cal F}(a,T)-\frac{k_BT}{4a^3}(1-r_0)\alpha(0),
\label{eq10}
\end{equation}
\noindent
where ${\cal F}(a,T)$ is presented in Eq.~(\ref{eq5}). Then the entropy at
zero temperature is obtained using the first equality in  Eq.~(\ref{eq6})
\begin{equation}
\tilde{S}(a,0)=\frac{k_B}{4a^3}(1-r_0)\alpha(0)>0.
\label{eq11}
\end{equation}
\noindent
This is in violation of the third law of thermodynamics because the
Casimir-Polder entropy at $T=0$ depends on the parameters of the system,
such as separation distance, static dielectric permittivity and static
atomic polarizability. In Ref.~\refcite{12} the contradiction between the
Lifshitz theory of atom-wall interaction with included dc conductivity of the
wall and thermodynamics is explained by the violation of thermal
equilibrium. The drift current of charge carriers
$\mbox{\boldmath$j$}=\sigma\mbox{\boldmath$E$}$ incorporated in the
permittivity (\ref{eq8}) leads to Joule's heating of the wall (Ohmic
losses).\cite{19} Then to preserve the temperature constant, one should
admit the existence of an unidirectional flux of heat from the wall
to the heat bath.\cite{20} Such fluxes are excluded in the state of
thermal equilibrium. In fact the drift current is irreversible process
which goes with an increase of entropy and brings a system out of thermal
equilibrium. Thus, it cannot be included in the Lifshitz theory formulated
for equilibrium systems.

The low-temperature behavior of the Casimir-Polder free energy and entropy
for an atom (molecule) near a metal wall was considered in Ref.~\refcite{6}.
Metal of the wall was described by means of the plasma model
\begin{equation}
\tilde\varepsilon_l^p=1+\frac{\omega_p^2}{\omega_c^2\zeta_l^2},
\label{eq12}
\end{equation}
\noindent
where $\omega_p=(4\pi e^2n/m)^{1/2}$ is the plasma frequency,
$n$ is the density of charge carriers,
 $e$ is the charge of an electron and $m$ is its
effective mass.
Similar results can be also obtained using the Drude model for the description
of the metal.\cite{22} The reason is that for metals not the TM reflection
coefficient at $\zeta_0=0$, but the TE one leads to the discontinuity
in the zero-frequency contribution to the Lifshitz formula.\cite{23}
However, for an atom-wall interaction the TE reflection coefficient at zero
frequency in Eq.~(\ref{eq1}) is multiplied by $\zeta_0^2=0$ and thus does
not contribute to the result.

Under the conditions $\tau\ll 1$ and $\delta_0/a\equiv c/(a\omega_p)\ll 1$
the leading terms of the asymptotic expressions for the Casimir-Polder
free energy and entropy are given by\cite{6}
\begin{eqnarray}
&&
{\cal F}^p(a,T)\approx E^p(a)-\frac{\hbar c\pi^3}{360a^4}\alpha(0)
\left(\frac{T}{T_{\rm eff}}\right)^4,
\nonumber \\
&&
S^p(a,T)\approx
\frac{\pi^3k_B}{45a^3}\alpha(0)
\left(\frac{T}{T_{\rm eff}}\right)^3,
\label{eq13}
\end{eqnarray}
\noindent
where $E^p(a)$ is the energy of atom-wall interaction at $T=0$ calculated
using the plasma model. Note that the results (\ref{eq13}) are the same as for
an atom near an ideal metal plate. Small corrections to these results depending
on atomic and material properties ($\beta$ and $\delta_0$) are obtained in
Ref.~\refcite{6}. For us it is only important that the entropy in
Eq.~(\ref{eq13}) vanishes when $T\to 0$, i.e., in the case of a metallic
wall the Nernst heat theorem is satisfied.

\section{Experimental Test for the Influence of Charge Carriers}

Important experimental test for the role of dc conductivity in the
interaction of atoms with a dielectric wall was performed in Ref.~\refcite{5}.
This is the measurement of the thermal Casimir-Polder force through
center-of-mass oscillations of the trapped Bose-Einstein condensate.
In that experiment the dipole oscillations with the frequency $\omega_0$
were excited in a ${}^{87}$Rb Bose-Einstein condensate separated by a distance
of a few micrometers from a fused-silica (SiO${}_2$) wall.
The Casimir-Polder force acting between a Rb atom and a wall changed the
magnitude of the oscillation frequency making it equal to some
$\omega_z$. The fractional frequency difference
$\gamma_z=|\omega_0-\omega_z|/\omega_0$ was measured and compared with
theory at an environment temperature $T_E=310\,$K and at different wall
temperatures $T_W=310\,$K (thermal equilibrium) and $T_W=479\,$K, 605\,K
(out of thermal equilibrium).

In the original publication\cite{5} the wall material was considered as a
dielectric with finite static dielectric permittivity
$\varepsilon_0=3.81$. Under this assumption, computations using the
Lifshitz formula (\ref{eq1})   and respective expression for situations
out of thermal equilibrium\cite{25} demonstrated a very good agreement
between experiment and theory at a 70\% confidence level. However, as was
discussed in Sec.~2, at nonzero temperature SiO${}_2$ has nonzero
conductivity which is ionic in nature and varies from
$10^{-9}\,\mbox{s}^{-1}$ to $10^{2}\,\mbox{s}^{-1}$ depending on the
concentration of alkali ions which are always present as trace constituents.
In Ref.~\refcite{26} the fractional frequency difference $\gamma_z$ was
recalculated both in equilibrium and out of thermal equilibrium by taking
into account a nonzero dc conductivity of fused silica in accordance
to Eq.~(\ref{eq8}). It was shown that the inclusion of the dc conductivity
leads to drastically different theoretical results both in thermal
equilibrium and out of thermal equilibrium which are in disagreement with the
measurement data.

\begin{figure}[pb]
\vspace*{-11.cm}
\centerline{\hspace*{15mm}\psfig{file=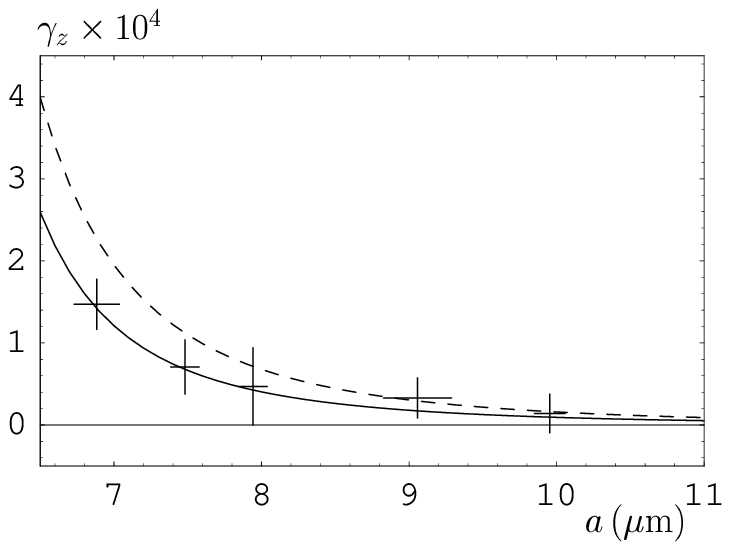,width=18cm}}
\vspace*{-9.9cm}
\caption{Fractional frequency difference versus separation in thermal
equilibrium with $T_W=T_E=310\,$K computed by neglecting (solid line) and
including (dashed line) the conductivity of the wall. The experimental data
are shown as crosses.
}
\end{figure}
In Fig.~1 the experimental data for $\gamma_z$ obtained\cite{5} in thermal
equilibrium are shown as crosses at different separations. The arms of
the crosses indicate the experimental errors determined individually
at each data point at a 70\% confidence level.\cite{5} The solid and dashed
lines show the theoretical results computed with neglect and inclusion of
the dc conductivity, respectively. Importantly, the obtained results with
included dc conductivity do not depend on its magnitude, but only on the
fact that it is nonzero. As is seen in Fig.~1, the first two experimental
points are in clear disagreement with the dashed line taking into account
the conductivity of fused silica. Similar conclusions are obtained in
situations out of thermal equilibrium. In Fig.~2(a,b) the same information,
as in Fig.~1, is provided for wall temperatures $T_W=479\,$K and 605\,K,
respectively. As is seen in Fig.~2, in nonequilibrium situations the
disagreement between the experimental data shown as crosses and the theory
taking into account the dc conductivity of a SiO${}_2$ wall (the dashed lines)
further widens.
\begin{figure}[pt]
\vspace*{-14.8cm}
\centerline{\hspace*{27mm}\psfig{file=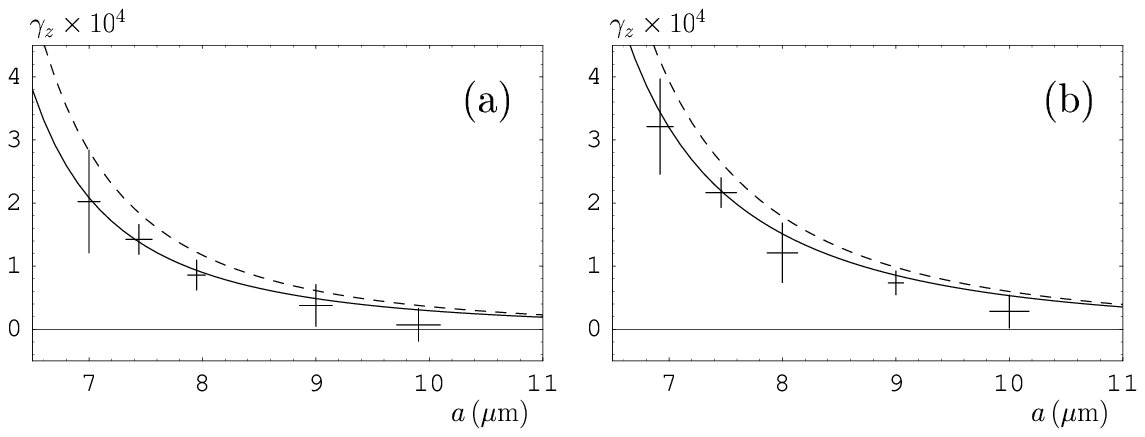,width=21cm}}
\vspace*{-10.6cm}
\caption{Fractional frequency difference versus separation out of thermal
equilibrium with (a) $T_W=479\,$K, $T_E=310\,$K and (b) $T_W=605\,$K,
$T_E=310\,$K. The other notations are the same as in Fig.~1.
}
\end{figure}
As is seen in Fig.~2(a), the three experimental points exclude the dashed
line and the other two only touch it. In Fig.~2(b) all data points
exclude the theoretical prediction incorporating the dc conductivity
of a SiO${}_2$ wall.\cite{26} The solid lines in Fig.~2(a,b) computed with
the dc conductivity neglected are in a very good agreement with the
measurement data. Thus, the inclusion of the dc conductivity of
dielectric materials into the Lifshitz theory is not only in contradiction
with thermodynamics (see Sec.~2), but is also experimentally inconsistent.

\section{Attempts to Include the Screening Effects}

Recently several attempts were undertaken\cite{8}\cdash\cite{10} to solve
the problem of free charge carriers in the Lifshitz theory through the
inclusion into consideration of screening effects and diffusion currents.
Here, we discuss these attempts only in the case of atom-wall interaction
(for the case of plate-plate interaction see Ref.~\refcite{27}).
The consideration of the scattering problem with account of both the
drift and diffusion currents of free charge carriers through use of
the Boltzmann transport equation results in the modified reflection
coefficients for the TM and TE modes of the electromagnetic field.\cite{9}

The modified TM reflection coefficient is given by
\begin{equation}
{r}_{\rm TM}^{\rm mod}({\rm i}\zeta_l,y)=\frac{\tilde\varepsilon_l y
-\bigl[y^2+(\tilde\varepsilon_l-1)\zeta_l^2\bigr]^{1/2}-
(y^2-\zeta_l^2)(\tilde\varepsilon_l-
\varepsilon_l)\tilde\eta_l^{-1}
\varepsilon_l^{-1}}{\tilde\varepsilon_l y
+\bigl[y^2+(\tilde\varepsilon_l-1)\zeta_l^2\bigr]^{1/2}+
(y^2-\zeta_l^2)(\tilde\varepsilon_l-
\varepsilon_l)\tilde\eta_l^{-1}
\varepsilon_l^{-1}},
\label{eq14}
\end{equation}
\noindent
where $\varepsilon_l,\>\tilde\varepsilon_l$ are defined in Eqs.~(\ref{eq3}),
(\ref{eq8}) and
\begin{equation}
\tilde\eta_l=\left[y^2-\zeta_l^2+\kappa_a^2\frac{\varepsilon_0
\tilde\varepsilon_l}{\varepsilon_l
(\tilde\varepsilon_l-\varepsilon_l)}
\right]^{1/2}, \qquad
\kappa_a\equiv 2a\kappa.
\label{eq15}
\end{equation}
\noindent
The parameter $\kappa$ in Eq.~(\ref{eq15}) is the inverse of the screening
length. If the density of charge carriers $n$ is sufficiently small, so
that they are described by the classical Maxwell-Boltzmann statistics,
one gets the Debye-H\"{u}ckel screening length\cite{28}
\begin{equation}
R_{\rm DH}=\frac{1}{\kappa_{\rm DH}}=
\sqrt{\frac{\varepsilon_0k_BT}{4\pi e^2n}}.
\label{eq16}
\end{equation}
\noindent
Assuming high density of charge carriers obeying the Fermi-Dirac statistics,
one arrives at the Thomas-Fermi screening length\cite{28}
\begin{equation}
R_{\rm TF}=\frac{1}{\kappa_{\rm TF}}=
\sqrt{\frac{\varepsilon_0E_F}{6\pi e^2n}},
\label{eq17}
\end{equation}
\noindent
where $E_F=\hbar\omega_p$ is the Fermi energy. As to the modified TE
reflection coefficient, ${r}_{\rm TE}^{\rm mod}({\rm i}\zeta_l,y)$,
it is given by the standard Eq.~(\ref{eq2}) where the permittivity
$\varepsilon_l$ is replaced with $\tilde\varepsilon_l$.

At zero Matsubara frequency the modified reflection coefficients are
given by
\begin{equation}
{r}_{\rm TM}^{\rm mod}(0,y)=\frac{\varepsilon_0\sqrt{y^2+\kappa_a^2}-
y}{\varepsilon_0\sqrt{y^2+\kappa_a^2}+y},
\qquad
{r}_{\rm TE}^{\rm mod}(0,y)=0.
\label{eq18}
\end{equation}
\noindent
The coefficient ${r}_{\rm TM}^{\rm mod}(0,y)$ from Eq.~(\ref{eq18}) was
first obtained in Ref.~\refcite{8}, where only the static case and
atom-wall interaction at large separations were considered. In so doing, all
other reflection coefficients were assumed to be unmodified.
The reflection coefficients (\ref{eq18}) were also reobtained in
Ref.~\refcite{10} using the phenomenological nonlocal approach.
At all nonzero Matsubara frequencies the coefficients in Refs.~\refcite{9}
and \refcite{10} differ between themselves and from unmodified
coefficients only slightly. As shown in Ref.~\refcite{27}, in application
to the case of two parallel plates the substitution of the modified
reflection coefficients into the Lifshitz formula results in violation of
the Nernst heat theorem for a wide class of materials and is excluded
experimentally. Here, we perform the thermodynamic test of the modified
reflection coefficients for atom-wall interaction at all Matsubara
frequencies (in Ref.~\refcite{12} only the approach of Ref.~\refcite{8}
was tested where the reflection coefficients at all nonzero Matsubara
frequencies remained unmodified).

References~\refcite{8}--\refcite{10} substitute the modified reflection
coefficients into the standard Lifshitz formula (\ref{eq1}) for the free
energy of dispersion interaction. We start with the case of dielectric wall
and find the asymptotic behavior of the free energy at low temperature.
For this purpose we introduce the small parameter
\begin{equation}
\beta_l=\frac{4\pi\sigma(0,T)}{\omega_c\zeta_l}\ll1,
\qquad l\geq 1,
\label{eq19}
\end{equation}
\noindent
which goes to zero due to Eq.~(\ref{eq7}) when temperature vanishes.
Then we expand the modified reflection coefficients
$r_{\rm TM,TE}^{\rm mod}({\rm i}\zeta_l,y)$ with $l\geq 1$ in powers
of this parameter
\begin{eqnarray}
&&
{r}_{\rm TM}^{\rm mod}({\rm i}\zeta_l,y)={r}_{\rm TM}({\rm i}\zeta_l,y)+
\beta_l\frac{\varepsilon_l y[2y^2+(\varepsilon_l-2)\zeta_l^2]}{\sqrt{y^2+
(\varepsilon_l-1)\zeta_l^2}[\varepsilon_l y+\sqrt{y^2+
(\varepsilon_l-1)\zeta_l^2}]^2} +O(\beta_l^2),
\nonumber \\[-2mm]
&& \label{eq20}\\[-2mm]
&&
{r}_{\rm TE}^{\rm mod}({\rm i}\zeta_l,y)={r}_{\rm TE}({\rm i}\zeta_l,y)+
\beta_l\frac{y[y-\sqrt{y^2+(\varepsilon_l-1)\zeta_l^2}]}{\sqrt{y^2+
(\varepsilon_l-1)\zeta_l^2}[ y+\sqrt{y^2+
(\varepsilon_l-1)\zeta_l^2}]} +O(\beta_l^2).
\nonumber
\end{eqnarray}
\noindent
Here, the reflection coefficients $r_{\rm TM,TE}$ are defined by
Eq.~(\ref{eq2}) with the dielectric permittivity of core electrons (\ref{eq3}).
The Casimir-Polder free energy ${\cal F}(a,T)$ calculated with these
 coefficients vanishes with temperature as $\sim\!(T/T_{\rm eff})^4$
according to Eq.~(\ref{eq5}). Substituting (\ref{eq20}) into the free
energy of atom-wall interaction (\ref{eq1}) one obtains
\begin{equation}
{\cal F}^{\rm mod}(a,T)={\cal F}(a,T)-\frac{k_BT\alpha(0)}{8 a^3}\left[
\int_{0}^{\infty}\!\!\!y^2dy\,{r}_{\rm TM}^{\rm mod}(0,y){\rm e}^{-y}
-2r_0+Q_1(T)\right],
\label{eq21}
\end{equation}
\noindent
where $Q_1(T)$ vanishes exponentially when $T\to 0$.
Calculating the negative derivative of both sides of (\ref{eq21})
with respect to $T$,
we arrive at
\begin{eqnarray}
&&
{S}^{\rm mod}(a,T)=S(a,T)+\frac{k_B\alpha(0)}{8 a^3}\left[
\vphantom{\int\limits_{0}^{0}}
\int_{0}^{\infty}\!\!\!y^2dy\,{r}_{\rm TM}^{\rm mod}(0,y){\rm e}^{-y}
-2r_0
\right.
\label{eq22} \\
&&~~~\left.
+\varepsilon_0T\frac{\partial\kappa_a^2}{\partial T}
\int_{0}^{\infty}\!\!\!dy\frac{y^3{\rm e}^{-y}}{\sqrt{y^2+
\kappa_a^2}(\varepsilon_0
\sqrt{y^2+\kappa_a^2}+y)^2}+Q_1(T)+TQ_1^{\prime}(T)
\right],
\nonumber
\end{eqnarray}
\noindent
where $S(a,T)$ is defined in Eq.~(\ref{eq6}) and
$\kappa_a=2a\kappa_{\rm DH}$.

Now we are in a position to perform the thermodynamic test for the approaches
taking the screening effects into account. The dc conductivity can be
represented as
\begin{equation}
\sigma(0,T)=\mu(T)\,|e|\,n(T),
\label{eq23}
\end{equation}
\noindent
where $\mu(T)$ is the mobility of charge carriers. For dielectric material
with exponentially decaying $n(T)$ in the limit of low $T$,
${r}_{\rm TM}^{\rm mod}(0,y)\to r_0$ and all terms on the right-hand side
of Eq.~(\ref{eq22}) added to $S(a,T)$ go to zero when $T\to 0$.
As a result, $S^{\rm mod}(a,T)$ goes to zero when $T$ vanishes, following
the same law, $\sim\!(T/T_{\rm eff})^3$, as $S(a,T)$, and the Nernst heat
theorem is satisfied. However, for many dielectric materials (such as
semiconductors doped below critical dopant concentration,
semimetals of dielectric type, some amorphous
semiconductors  or solids with
ionic conductivity) $n(T)$ does not go to zero in the limit $T\to 0$.
For such materials the dc conductivity $\sigma(0,T)$ goes to zero
exponentially fast (as for all dielectrics) due to the vanishing
mobility $\mu(T)$ in Eq.~(\ref{eq23}). For all dielectric materials
with nonvanishing $n$ it holds ${r}_{\rm TM}^{\rm mod}(0,y)\to 1$ when
$T$ and $\mu(T)$ simultaneously vanish. This is because
$\kappa_{\rm DH}\to\infty$ when $T\to 0$ in accordance with Eq.~(\ref{eq16}).
In this case Eq.~(\ref{eq22}) results in
\begin{equation}
S^{\rm mod}(a,0)=\frac{k_B}{4a^3}\,(1-r_0)\,\alpha(0)>0,
\label{eq24}
\end{equation}
\noindent
i.e., the Nernst heat theorem is violated in the same way as in the
standard Lifshitz theory with included dc conductivity
[see  Eq.~(\ref{eq11})]. Mathematically, the reason of violation is the
discontinuity of the TM reflection coefficient as a function of $\xi$
and $T$ at the origin of the $(\xi,T)$-plane.\cite{29}
Note that claim of Ref.~\refcite{10} that the approach taking the
screening effects into account is consistent with thermodynamics is
incorrect. As shown in Refs.~\refcite{27},\,\refcite{30}, this claim is based
on misinterpretation of relevant physical quantities.
Similar claim made in Ref.~\refcite{28a} is also incorrect.
The proof of the validity of the Nernst theorem given
in Ref.~\refcite{28a} uses the condition that $n\to 0$ when
$T$ vanishes. Thus, this proof simply ignores wide classes of
dielectric materials for which this is not so and the Nernst
theorem is violated.

The physical reason for the inconsistency with thermodynamics discussed
above is the violation of the applicability conditions of the Lifshitz theory.
The drift and diffusion currents described by the Boltzmann transport
equation are irreversible processes out of thermal equilibrium which go
with an increase of the entropy. As to the Lifshitz theory, it is
formulated under the condition of thermal equilibrium and describes
equilibrium systems. Because of this, the substitution of the modified
reflection coefficients into the standard Lifshitz formula (\ref{eq1})
leads to thermodynamic puzzles.

In the end of this section we consider atom-wall interaction in the case
of metallic walls when the screening effects are taken into account.
In this case one should use $\kappa=\kappa_{\rm TF}$. For metals,
$\kappa_a=2a\kappa_{\rm TF}$ is very large and the inverse quantity
$\beta_a\equiv 1/\kappa_a\ll 1$ can be used as a small parameter.
The expansion of the TM reflection coefficient (\ref{eq14}) in powers of
$\beta_a$ takes the form
\begin{eqnarray}
&&
{r}_{\rm TM}^{\rm mod}({\rm i}\zeta_l,y)=\tilde{r}_{\rm TM}({\rm i}\zeta_l,y)-
2\beta_a\,Z_l+O(\beta_a^2),
\label{eq25} \\
&&
Z_l\equiv\sqrt{\frac{\tilde\varepsilon_l
(\tilde\varepsilon_l-\varepsilon_l)^3}{\varepsilon_0\varepsilon_l}}
\,\frac{y(y^2-\zeta_l^2)}{[\tilde\varepsilon_l y+\sqrt{y^2+
(\tilde\varepsilon_l -1)\zeta_l^2}]^2}.
\nonumber
\end{eqnarray}
\noindent
Here, $\tilde{r}_{\rm TM}({\rm i}\zeta_l,y)$ is defined by Eq.~(\ref{eq2})
where $\varepsilon_l$ is replaced with $\tilde\varepsilon_l$ given in
Eq.~(\ref{eq8}). After the substitution of Eq.~(\ref{eq25}) and the
expression for ${r}_{\rm TE}({\rm i}\zeta_l,y)$ (the latter coincides
with the known expression for the TE reflection coefficient\cite{31}
obtained using the Drude model) into Eq.~(\ref{eq1}), one obtains
\begin{equation}
{\cal F}^{\rm mod}(a,T)={\cal F}^{p}(a,T)+\beta_a{\cal F}^{(\beta)}(a,T),
\label{eq26}
\end{equation}
\noindent
where ${\cal F}^{p}(a,T)$ is defined in Eq.~(\ref{eq13}) and the quantity
${\cal F}^{(\beta)}(a,T)$ originates from the second contribution on the
right-hand side of Eq.~(\ref{eq25}). It is easily seen that at low $T$
the quantity ${\cal F}^{(\beta)}$ behaves as
${\cal F}^{(\beta)}(a,T)=E^{\beta)}(a)+O(T^5/T_{\rm eff}^5)$.
Calculating the negative derivative with respect to temperature from
both sides of Eq.~(\ref{eq26}), we arrive to the equality
$S^{\rm mod}(a,0)=0$, i.e., in the case of metallic walls the Nernst heat
theorem is satisfied. This result is preserved if metal wall is described
by means of the Drude model.

\section{Conclusions and Discussion}

In the foregoing we have discussed problems in the Lifshitz theory of
atom-wall interaction connected with the description of free charge
carriers in wall material. It was shown that in the case of metallic
walls this does not create any problem. As to dielectric walls,
thermodynamically and experimentally consistent results are obtained when
the dc conductivity of wall material is neglected. The inclusion of the
dc conductivity into the model of dielectric response results in the
violation of the Nernst heat theorem and in contradictions with
experimental data. Similar situation also holds in recently proposed
theoretical approaches attempting to include in the Lifshitz theory the
screening effects and diffusion currents. For metallic walls this does not
create problems, but leads to contradictions with thermodynamics for wide
classes of dielectrics. Consistent theoretical description of atom-wall
interaction in the case of dielectric walls uses the rule that the dc
conductivity of dielectric material should be neglected.\cite{26}

This rule should be considered  as a phenomenological one, but it can
be justified in terms of most basic physical concepts on the subject.
The point is that the physical phenomenon of dispersion forces occurs in
thermal equilibrium and is caused by the fluctuating electromagnetic
field. The latter cannot transmit energy in only one direction from the
field to charge carriers, heat the wall material and create
in homogeneous media the nonhomogeneous
concentrations of charges resulting in the screening and diffusion effects.
All these phenomena are of irreversible character, lead to an increase
of the entropy and can occur in external electric field. Thus, we get
the conclusion that there is a fundamental difference between the
fluctuating electromagnetic field (including its ``static component''\cite{8})
and external electric field. The discussed above problems in both the
standard formulation of the Lifshitz theory of atom-wall interaction
and its generalization including the screening effects arise when the
properties of external field are attributed to the fluctuating field in
an unjustified way. Final clarification of this issue is expected in near
future.

\section*{Acknowledgments}

G.L.K. and V.M.M.\ are grateful to the Center of Theoretical Studies
and the Institute for Theoretical Physics, Leipzig University, for their
kind hospitality.
They were partially supported by
Deutsche Forschungsgemeinschaft, Grant No.~436\,RUS\,113/789/0--4.


\end{document}